**Enhancing Organizational Performance: Harnessing AI and NLP for User Feedback Analysis in Product Development**

Tian Tian Stuart Business School of Illinois Institute of Technology
Liu Zehui Veritas Collegiate Academy
Zichen Huang Shenzhen Middle School of Mathematics and Physics
Yubing Tang Stuart Business School of Illinois Institute of Technology

**Abstract**

This paper explores the application of AI and NLP techniques for user feedback analysis in the context of heavy machine crane products. By leveraging AI and NLP, organizations can gain insights into customer perceptions, improve product development, enhance satisfaction and loyalty, inform decision-making, and gain a competitive advantage. The paper highlights the impact of user feedback analysis on organizational performance and emphasizes the reasons for using AI and NLP, including scalability, objectivity, improved accuracy, increased insights, and time savings. The methodology involves data collection, cleaning, text and rating analysis, interpretation, and feedback implementation. Results include sentiment analysis, word cloud visualizations, and radar charts comparing product attributes. These findings provide valuable information for understanding customer sentiment, identifying improvement areas, and making data-driven decisions to enhance the customer experience. In conclusion, promising AI and NLP techniques in user feedback analysis offer organizations a powerful tool to understand customers, improve product development, increase satisfaction, and drive business success.

*Key words*: Artificial Intelligence (AI), Natural Language Processing (NLP), organizational performance, user feedback

## 1. Introduction

User feedback analysis is the process of collecting, organizing, and analyzing feedback from users of a product or service (De Vreede & Briggs, 2019). This feedback can come from various sources such as surveys, customer support tickets, social media posts, or direct comments from users (Leinonen, Hämäläinen, & Juntti, 2009; Bistritz & Leshem, 2018; Francis, Mehta, & Ananya, 2016). The purpose of user feedback analysis is to understand the experiences and perspectives of users, identify areas for improvement, and inform decision-making related to product development and customer experience (Ali, Hong, & Chung, 2021; Dąbrowski et al., 2019). Effective user feedback analysis based on Martikainen's research can help organizations to better understand their customers, improve their products and services, and ultimately drive business success (Kujala, 2008; Bragge, Merisalo-Rantanen, & Hallikainen, 2005; Cvitanovic et al., 2015).

### 1.1 Influences on Organizational Performance

Analysis of user feedback using AI and NLP techniques can bring significant meaning to organizational behavior in several ways:

**Identifying customer preferences and needs**: By analyzing user feedback, organizations can gain insights into how users perceive their products or services (Gensler et al., 2015). This analysis can help identify the features and aspects of a product that users value the most, allowing organizations to better understand customer preferences and needs. This knowledge can guide decision-making processes, such as product development or marketing strategies, enabling organizations to align their offerings with customer expectations (Intezari & Gressel, 2017).

**Improving product development and innovation**: User feedback analysis can provide valuable information for product development and innovation ((Liu et al., 2020). By understanding how users perceive a product or service, organizations can identify areas for improvement, uncover potential issues, and gather ideas for new features or enhancements (He et al., 2015). This analysis can guide the iterative design process, helping organizations build better products that cater to user demands and expectations.

**Enhancing customer satisfaction and loyalty**: User feedback analysis allows organizations to address customer concerns and issues promptly (Pomputius, 2019). By identifying recurring patterns or themes in feedback, organizations can identify areas where improvements are needed, enabling them to take proactive measures to address customer pain points (Manser, 2009, Zhang & Vásquez, 2014). This approach can enhance customer satisfaction and loyalty by demonstrating that the organization values and listens to its customers, fostering a positive relationship between the organization and its user base.

**Informing decision-making processes**: User feedback analysis can provide decision-makers with direct insights into user perspectives and sentiments (Elgendy & Elragal, 2014). By understanding how users perceive a product or service, organizations can make data-driven decisions based on customer feedback (Wang et al., 2022). This analysis can help validate or refute assumptions, support strategic planning, and guide resource allocation within the organization (McGrath, 2010). Ultimately, it allows decision-makers to align their actions with user expectations and drive organizational success.

**Competitive advantage and market differentiation**: User feedback analysis can provide organizations with a competitive advantage by uncovering unique insights about their products or services (Jusoh & Parnell, 2008). By leveraging AI and NLP techniques, organizations can gain a deep understanding of user perceptions, preferences, and expectations (Rathore, 2023). This knowledge can be used to differentiate themselves in the market by tailoring their offerings to meet customer needs more effectively than their competitors. It can also guide organizations in developing unique value propositions or positioning their products or services in a way that resonates with their target audience.

In summary, user feedback analysis can ultimately contribute to improving organizational behavior and performance. The paper constructed by 6 sections. Reasons of using Artificial Intelligence (AI) and natural language processing (NLP) to do user feedback analysis of heavy machine product are discussed in section 2. Section 3 is the method used of user feedback analysis. Section 4 summarized the contributions of this paper. Section 5 is limitations of the study. Lastly, section 6 presented the conclusion and future work.

## 2. AI and NLP in User Feedback Analysis of Heavy Machine Product

User feedback analysis using AI and natural language processing (NLP) involves using computational methods (Cambria & White, 2014; Balahur, Mihalcea, & Montoyo, 2014) and algorithms (Joseph et al., 2016) to analyze customer feedback data in text form (Patton et al., 2020). This can include reviews, surveys, support tickets, and other forms of customer communications. The goal of this analysis is to gain insights into customer sentiment, preferences, pain points, and satisfaction levels, which can then be used to improve products, services, and customer experiences. AI and NLP techniques can be used to identify key themes and patterns, categorize feedback into relevant topics, extract important information such as product features and customer attitudes, and even perform sentiment analysis (Patton et al., 2020) to determine the overall tone of the feedback.

There are several reasons why companies use AI and NLP to analyze user feedback:

**Scalability:** AI and NLP can process large amounts of text data much faster and more efficiently than humans, making it easier to analyze large volumes of customer feedback (Mills & Bourbakis, 2013).

**Objectivity**: AI algorithms can provide objective and consistent results, removing the subjectivity that can come with manual analysis.

**Improved accuracy**: AI and NLP models are trained on vast amounts of data (Baclic et al., 2020; Khosravi, Rouzrokh, & Erickson, 2022), which can lead to improved accuracy in sentiment analysis (Dhaoui, Webster, & Tan, 2017; Shakhovska, Shakhovska, & Veselý, 2020) and categorization of feedback compared to manual methods.

**Increased insights**: AI and NLP can identify patterns and trends (Pan & Zhang, 2021) in customer feedback that may not be immediately obvious to humans, providing valuable insights into customer preferences and pain points.

**Time savings**: By automating the analysis process, AI and NLP can save businesses time and resources that would otherwise be spent on manual analysis.

Therefore, the use of AI and NLP for user feedback analysis can help companies to make more informed decisions based on data-driven insights, leading to improved customer experiences and business outcomes.

## 3. Method

Text analysis and rating analysis are used in user feedback analysis here.

**Text analysis** involves the use of NLP techniques to process and analyze written customer feedback, such as reviews, support tickets, and surveys (Ravi & Ravi, 2015). This can include techniques such as sentiment analysis and topic modeling. The goal of text analysis is to extract insights and information from the unstructured data contained in customer feedback (Elgendy & Elragal, 2014).

**Rating analysis**, on the other hand, involves analyzing numerical ratings that customers provide, such as star ratings or numerical scores (Lak & Turetken, 2014). This type of analysis focuses on identifying patterns and trends in customer ratings, such as average ratings, distributions, and changes over time (Fu et al., 2013). The goal of rating analysis is to gain a quantitative understanding of customer satisfaction levels.

Both text analysis and rating analysis can be used together to provide a more comprehensive understanding of customer feedback. For example, sentiment analysis of customer reviews can be combined with rating analysis to determine whether the overall sentiment of customer feedback aligns with the numerical ratings provided.

This can help businesses to identify areas for improvement and make more informed decisions.

### 3.1 Steps of User Feedback Analysis

The method of user feedback analysis typically involves the following steps:

*Data Collection*: This involves gathering user feedback from questionnaire and direct comments from users. The data are collected in a consistent and organized manner to facilitate analysis. The research used three crane products, respectively, XCMG XCR100U, Tadano GR1000XL and Grove GRT8100.

Number of valid comments for each product are

XCMG: 76
Tadano: 49
Grove: 43

*Data Cleaning*: This involves removing any irrelevant or duplicate data and ensuring that the data is in a consistent format for analysis.

*Data Analysis*: This involves using various techniques to analyze the data, such as sentiment analysis, text classification, topic modeling, clustering, or regression analysis. The goal is to understand the experiences and perspectives of users, identify areas for improvement, and inform decision-making.

*Results Interpretation*: This involves interpreting the results of the data analysis to understand the key findings and insights and generate recommendations for action.

*Feedback Implementation*: This involves using the insights from the data analysis to make improvements to the product or service, enhance the customer experience, and drive business success.

### 3.2 Data Analysis

#### 3.2.1 Text Analysis

There are several text analysis methods are used to analyze user feedback, including:

Sentiment Analysis: This is a method that uses natural language processing (NLP) and machine learning algorithms to classify user feedback. Table 1 used mean and standard deviation to give the result of each product, and Figure. 1 listed positive (0.05+), negative

(below -0.05), and neutral (-0.05-0.05) in sentiment. Combined results in Table 1 and Fig. 1 show that:

The attitude towards XCMG XCR100U is neural.

The attitude towards Tadano GR1000XL is between neural and positive.

The attitude towards Grove GRT8100 is positive.

Overall, Grove GRT8100 has the best reviews.

Table. 1 Results of sentiment analysis of each product

| Brand | Mean | Std |
| --- | --- | --- |
| XCMG XCR100U | -0.017 | 0.331 |
| Tadano GR1000XL | 0.050 | 0.234 |
| Grove GRT8100 | 0.101 | 0.303 |

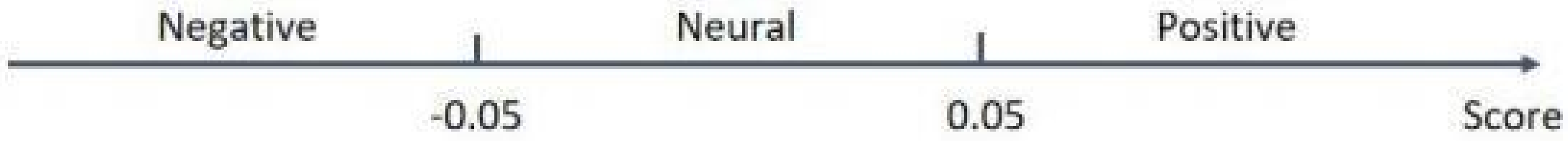

Figure. 1 Positive (0.05+), negative (below -0.05), and neutral (-0.05-0.05) in sentiment

Word Cloud: A word cloud is a visual representation of the most frequently occurring words in a given text. In the context of user feedback analysis, a word cloud can be used to visualize the most common themes and topics in customer feedback.

To create a word cloud, the text data is first pre-processed to remove stop words (such as "and" "the," "a," etc.), and the frequency of each remaining word is calculated. The words are then arranged on a graph with the size of each word proportional to its frequency. The result is a word cloud that provides a quick and intuitive view of the most prominent themes and topics in the customer feedback data.

Word clouds can be useful for quickly identifying common complaints, suggestions, and areas of interest in customer feedback. Fig. 2, 3, 4 are word cloud showed by visually representing XCMG XCR100U, Tadano GR1000XL, and Grove GRT8100. The word clouds can help businesses to focus their efforts on the areas that matter most to their customers.

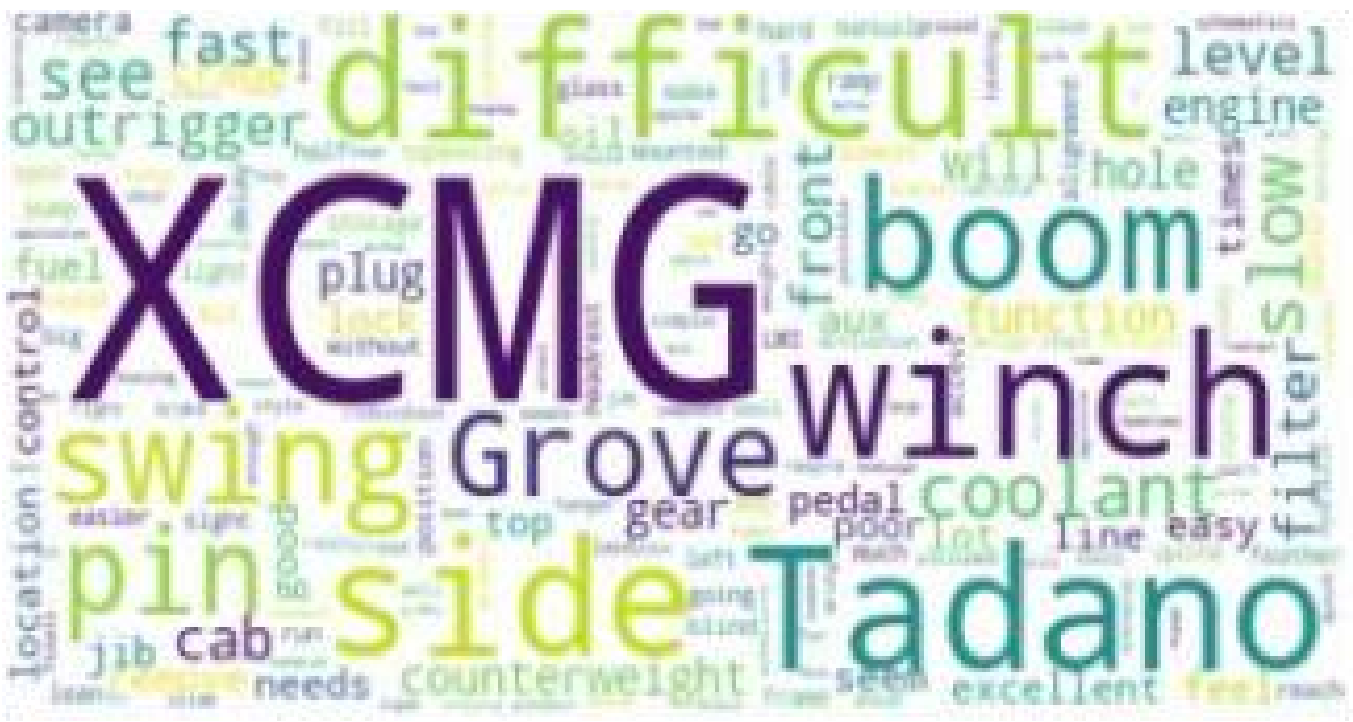

Fig. 2 Word cloud for XCMG XCR100U

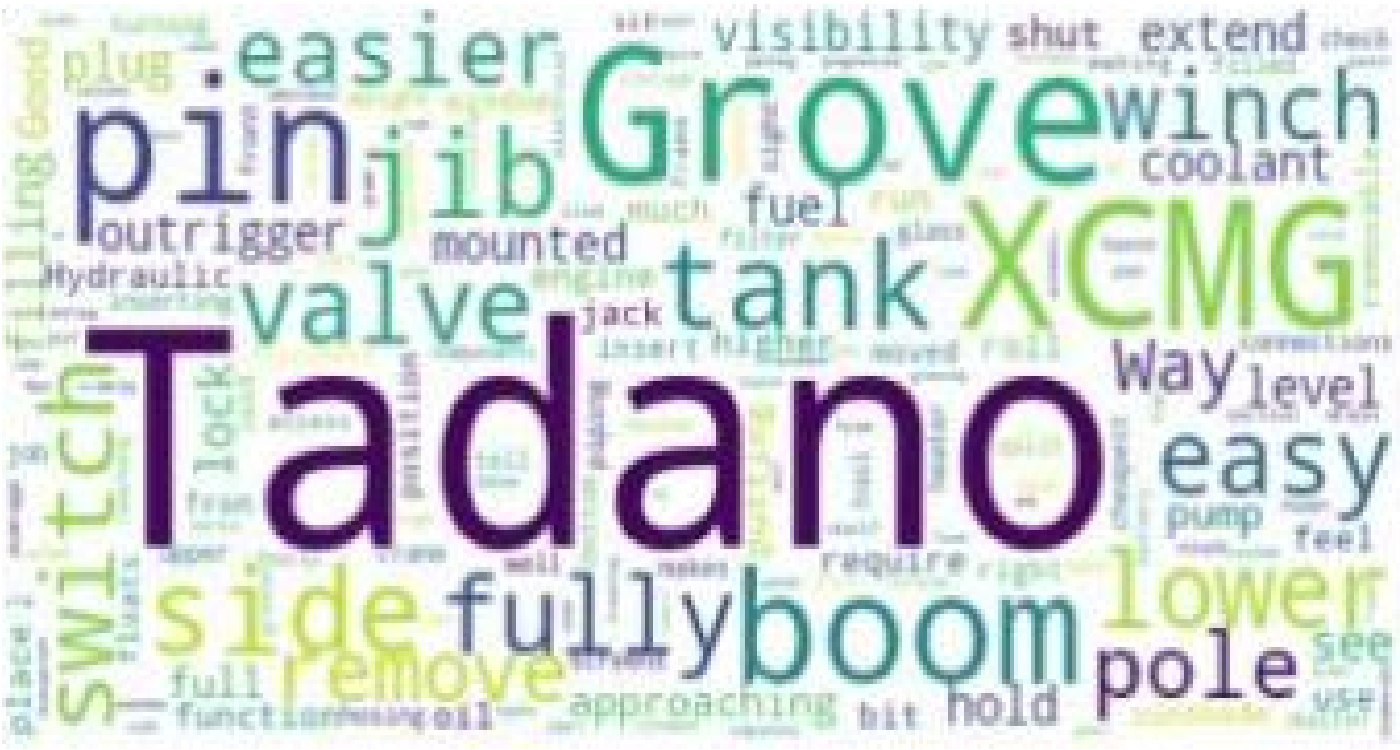

Fig. 3 Word cloud for Tadano GR1000XL

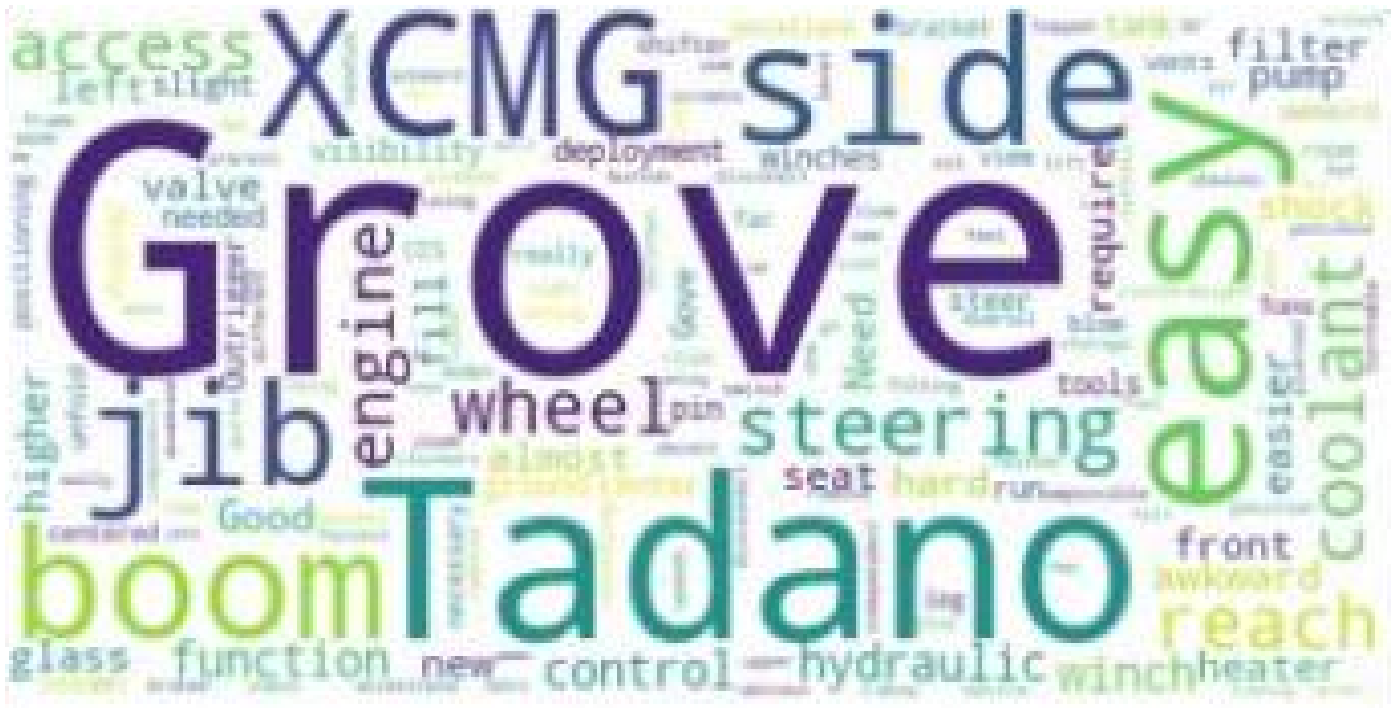

Fig. 4 World cloud of Grove GRT8100

Topic Modeling: This is a method that uses NLP techniques to identify the underlying topics or themes present in a large corpus of user feedback. This can help organizations to understand the key issues or concerns of their customers at a high level. This research designs 3 Topics of Comments for XCMG XCR100U.

Fig. 5 shows Top-20 most relevant terms for Topic 1 that is related to peripheral.

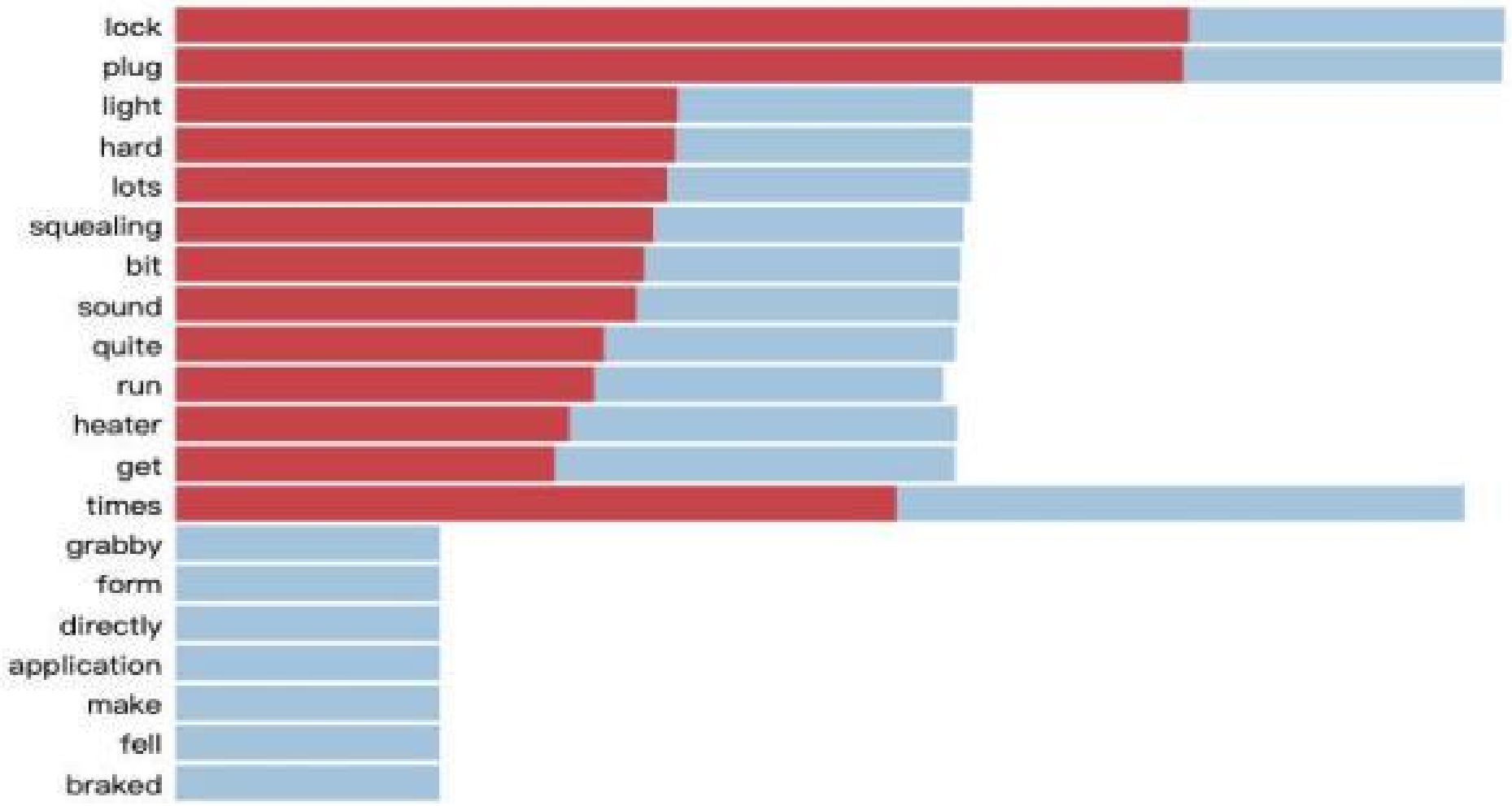

Fig. 5 Top-20 most relevant terms related to for Topic 1 peripherals

Fig. 6 shows Top-20 most relevant terms for Topic 2 that is related to functionality of XCMG XCR100U.

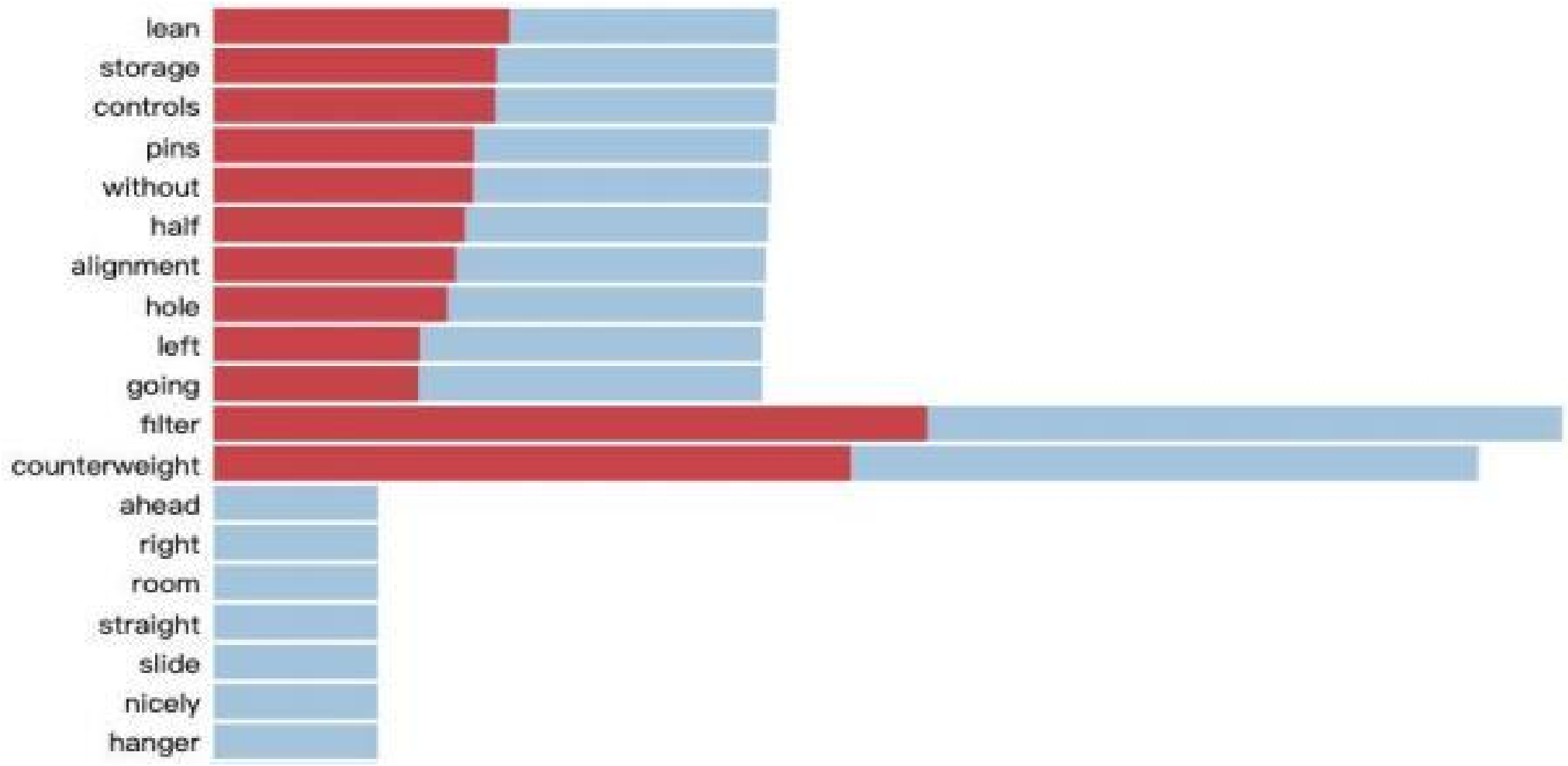

Fig. 6 Top-20 most relevant terms related to Topic 2 functions

Fig.7 shows the Top-20 most relevant terms for Topic 3 that is related to driving experience of XCMG XCR100U.

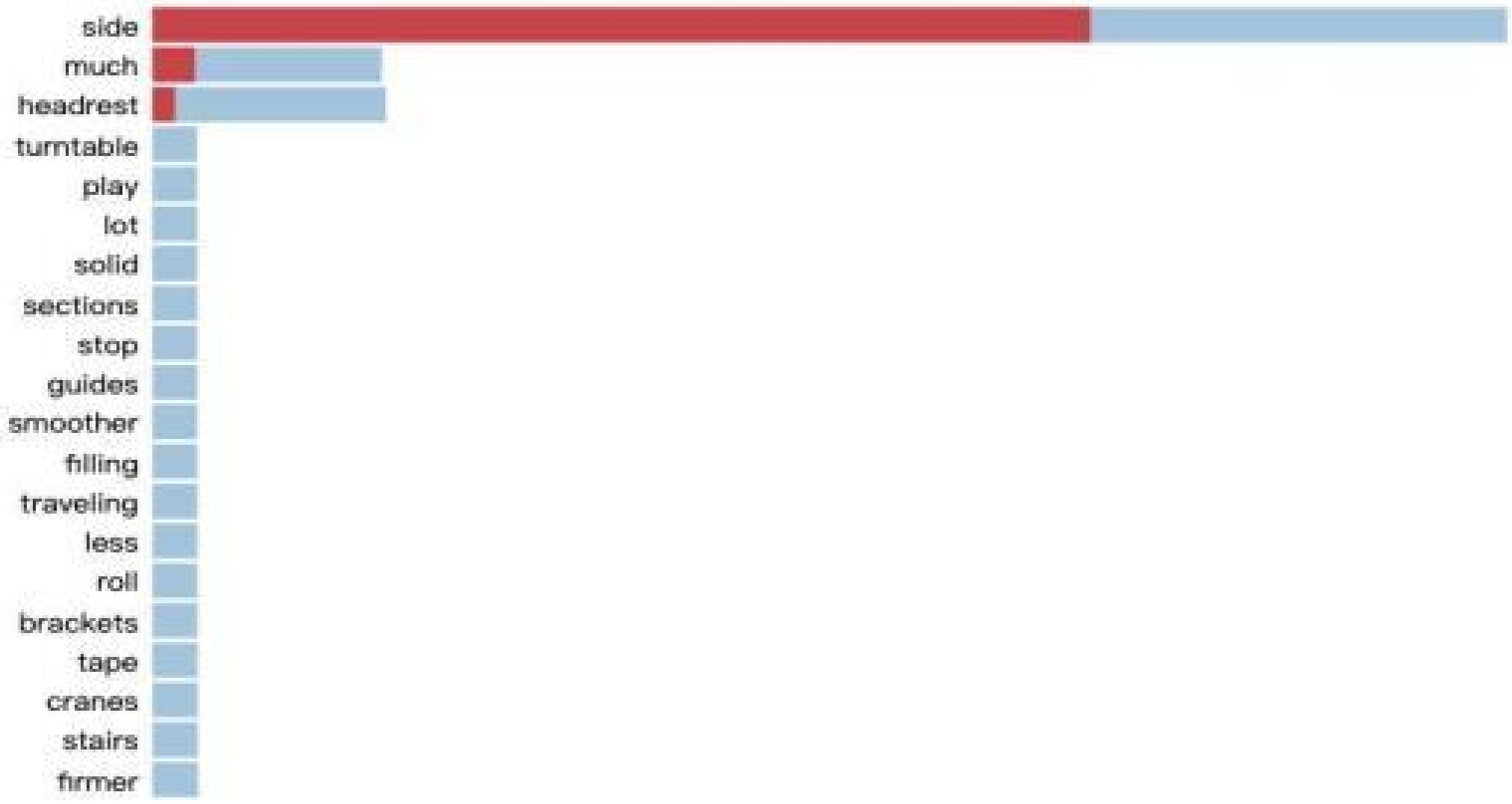

Fig.7 Top-20 most relevant terms related to Topic 3 driving experience

Analysis on Cab (Fig. 8) results as below:

XCMG XCR100U has advantage over space, seat comfort and HVAC aspects compared with competing products.

XCMG XCR100U has disadvantage over Visibility, Ease of operation and Accessibility aspects compared with competing products.

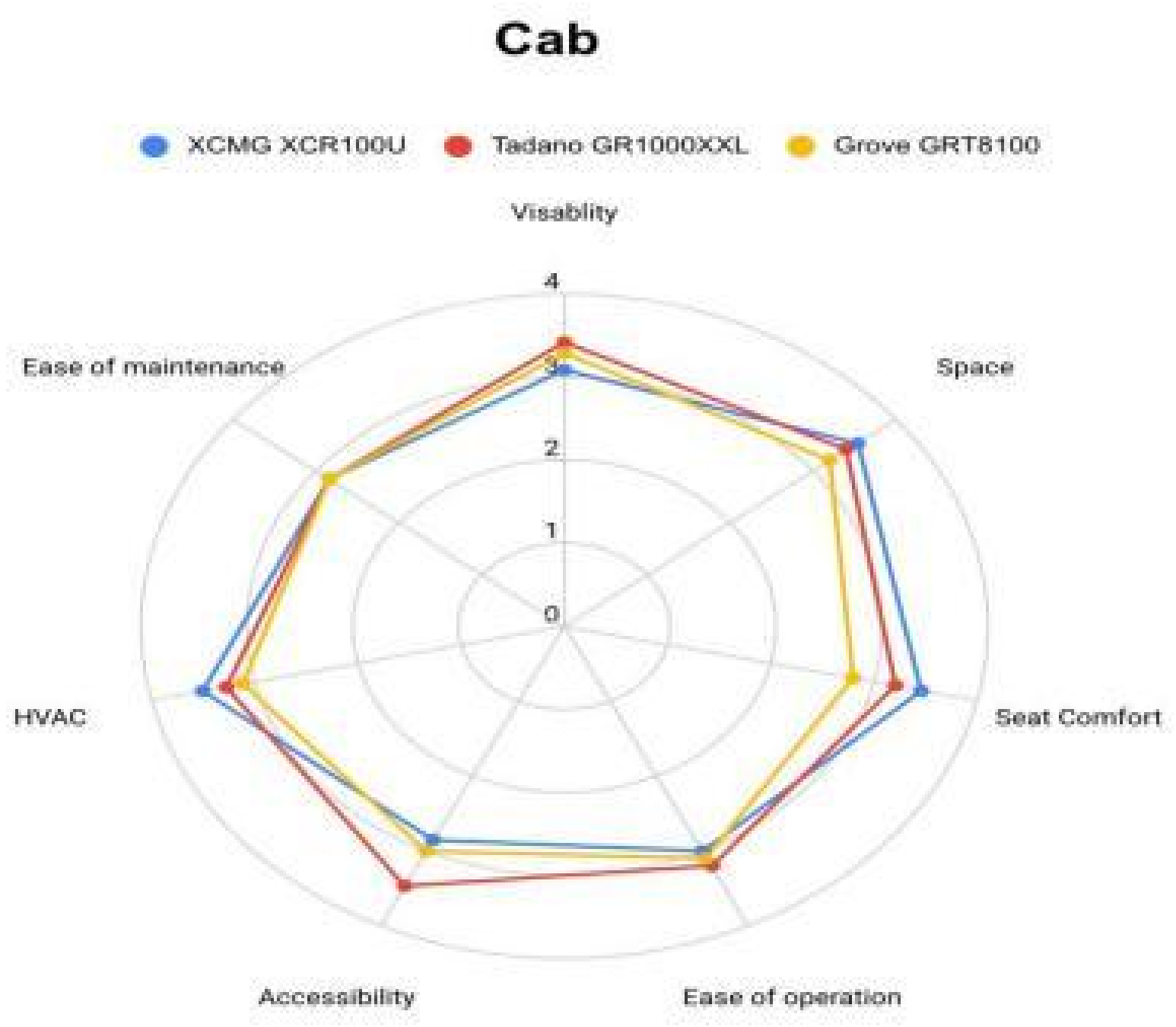

Fig. 8 Radar chart of Cab

Analysis on Systems for lifting operations (partially extended boom, rated load) (Fig. 9) results showed as below:

XCMG XCR100U has no obvious advantage compared with competing products.

XCMG XCR100U has disadvantage over Main winch and Slewing aspects compared with competing products.

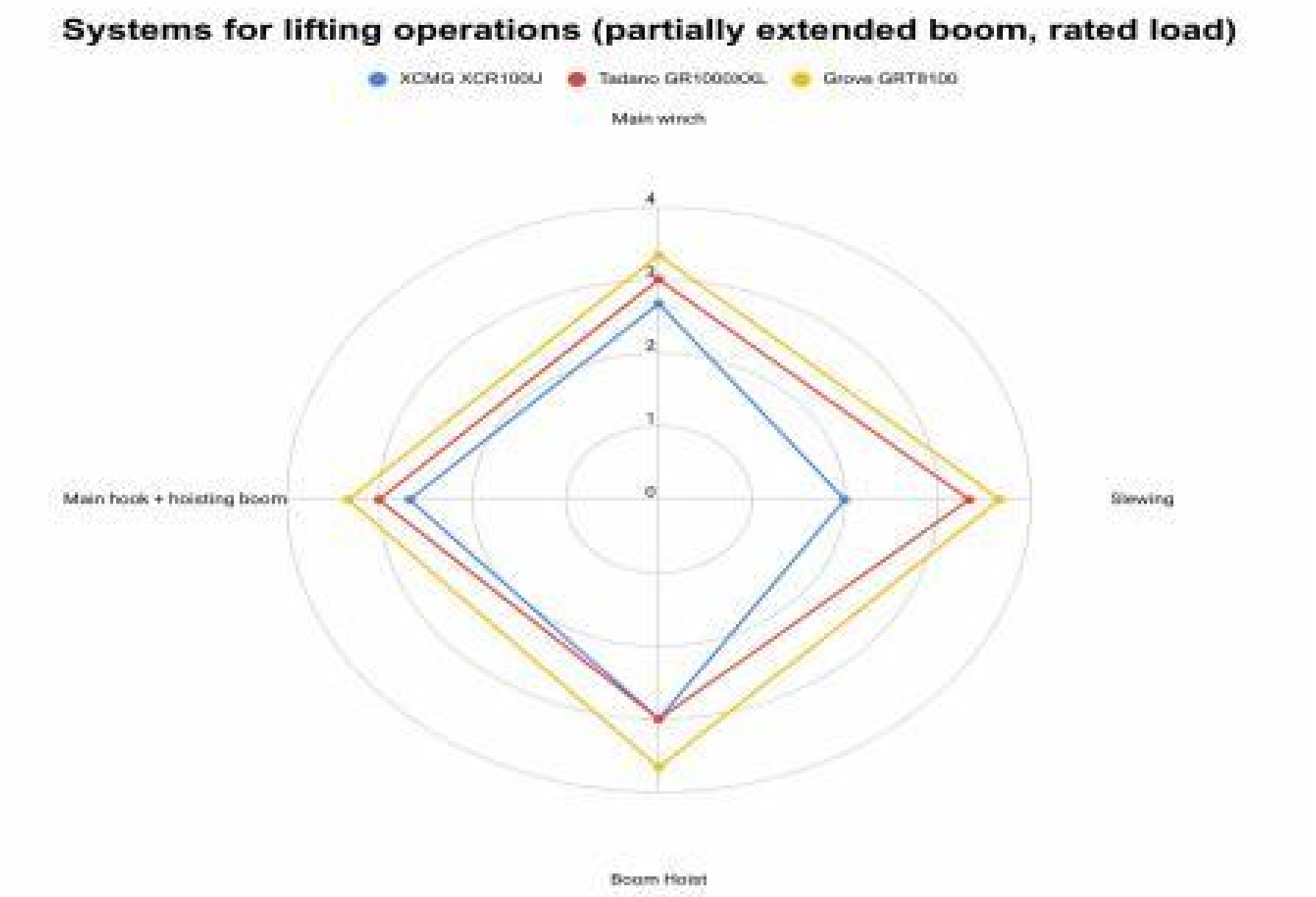

Fig. 9 Radar chat of systems for lifting operations (partially extended boom, rated load)

Analysis on Systems for lifting operations (no load) (Fig. 10) results showed as below:

XCMG XCR100U has no obvious advantage compared with competing products.

XCMG XCR100U has disadvantage over Main winch, Auxiliary winch and Slewing aspects compared with competing products.

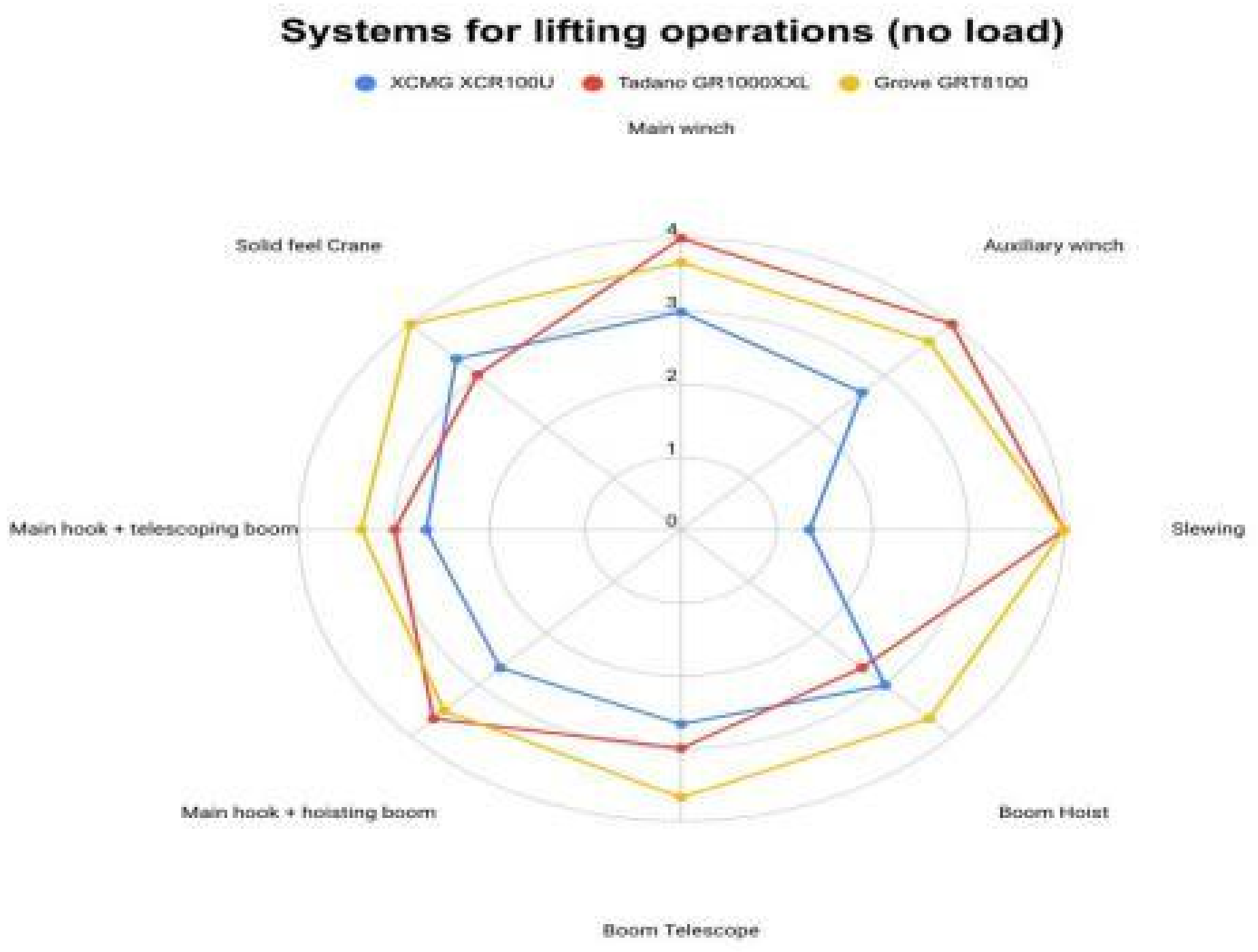

Fig. 10 Radar chat of systems for lifting operation (no load)

Analysis on Systems for lifting operations (partially extended boom, rated load) (Fig. 11) results showed as below:

XCMG XCR100U has no obvious advantage compared with competing products.

XCMG XCR100U has disadvantage over Main winch and Slewing aspects compared with competing products.

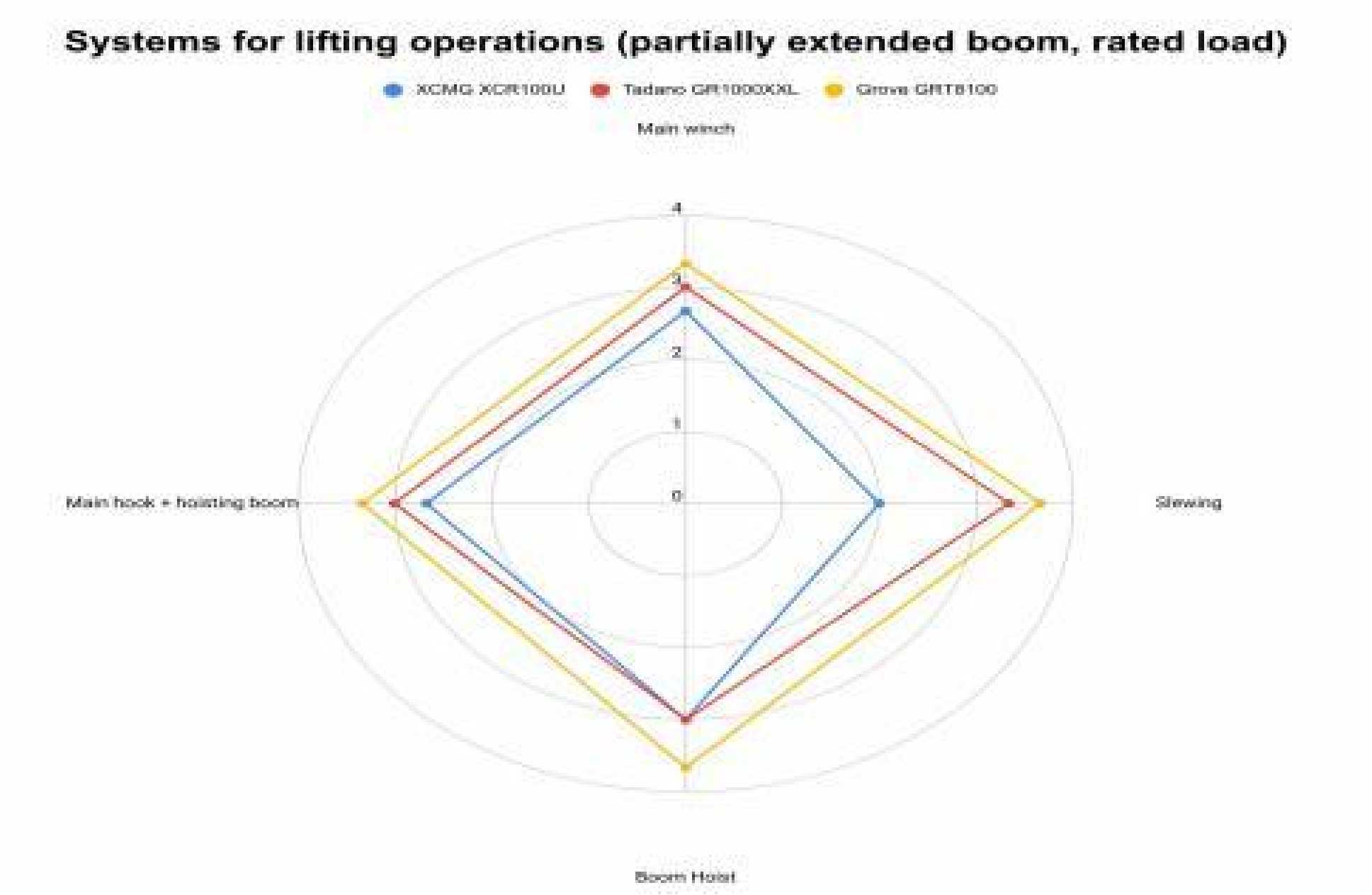

Fig. 11 Radar chart of systems for lifting operations (partially extended boom, rated load)

Analysis on Other systems (Fig. 12) results showed as below:

XCMG XCR100U has advantage over Cab tilting and Differential system aspects compared with competing products.

XCMG XCR100U has disadvantage over Suspension system aspect compared with competing products.

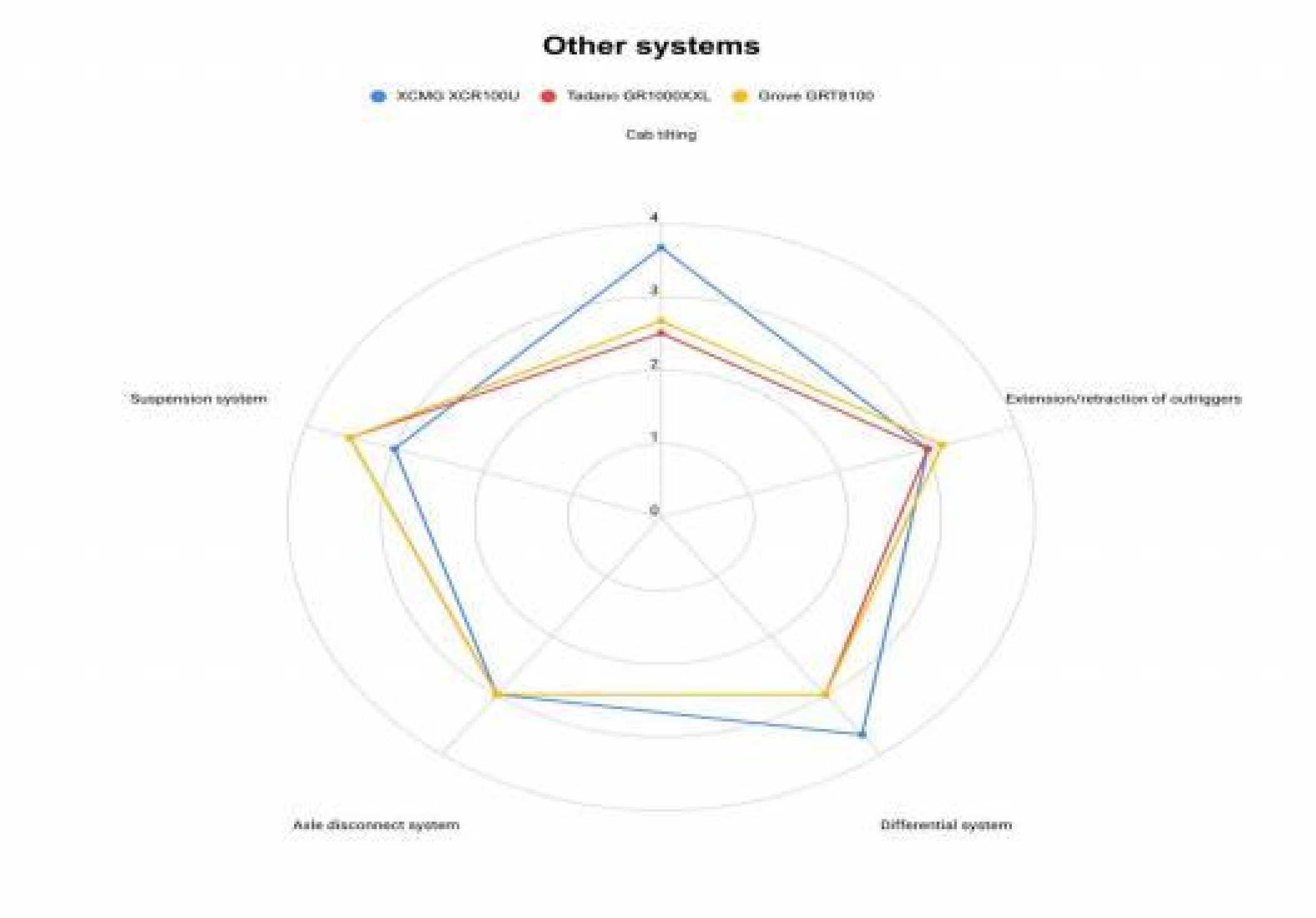

Fig. 12 Radar chart of other systems

To sum up, both text and rating analysis of AI methods can greatly enhance the efficiency and effectiveness of user feedback analysis, enabling organizations to make informed decisions based on a large amount of data. However, it's important to note that AI should be used as a tool to augment human analysis and decision-making, not replace it entirely.

## 4. Contributions

This paper contributes to the field of user feedback analysis in the context of heavy machine crane products by demonstrating the use of AI and NLP techniques. The main contributions of this paper are as follows:

**Practical Application**: The paper provides a practical example of how AI and NLP can be applied to analyze user feedback for heavy machine crane products. It demonstrates the potential of these techniques in gaining valuable insights into customer perceptions and preferences.

**Organizational Benefits**: The paper highlights the impact of user feedback analysis on organizational behavior and performance. It explains how AI and NLP can help organizations identify customer preferences and needs, improve product development and innovation, enhance customer satisfaction and loyalty, and inform decision- making processes. These benefits can ultimately lead to improved business outcomes and a competitive advantage.

**Methodological Approach**: The paper outlines a systematic method for user feedback analysis, including data collection, data cleaning, data analysis, results interpretation, and feedback implementation. It explains the use of text analysis and rating analysis techniques, such as sentiment analysis, word cloud visualization, and topic modeling, to extract insights from user feedback data.

**Specific Findings**: The paper presents specific findings from the user feedback analysis of three heavy machine crane products: XCMG XCR100U, Tadano GR1000XL, and Grove GRT8100. It includes the results of sentiment analysis, word cloud visualization, and rating analysis using radar charts. These findings provide valuable information about customer sentiment, common themes in feedback, and strengths and weaknesses of the products.

**Practical Implications**: The paper discusses the practical implications of the findings and offers recommendations for action. It suggests areas for improvement and highlights the unique value propositions of each product based on the user feedback analysis. These recommendations can guide organizations in making informed decisions to enhance their products, services, and customer experiences.

**Managerial Implications:** The findings of this study offer business insights for managers, suggesting that employing AI and NLP for user feedback analysis can enhance decision-making and customer relations. Through AI-enabled feedback analysis, managers can detect trends and customer needs that inform product development and market strategies. Additionally, sentiment analysis and topic modeling provide valuable data to refine and innovate product features in line with customer preferences. Implementing NLP for feedback processing also helps in promptly addressing customer issues, boosting engagement and satisfaction. Utilizing AI and NLP not only provides a competitive edge by deepening customer experience insights but also optimizes resource distribution towards crucial areas for customer retention. Managers should incorporate AI and NLP in operational processes to improve product quality, customer satisfaction, and financial outcomes.

### 4.1 Artificial Intelligence (AI) Contributions in User Feedback Analysis

Artificial Intelligence (AI) has made a significant contribution to user feedback analysis by automating the process and providing more accurate, consistent, and efficient results. Some of the key benefits of using AI for user feedback analysis include:

- Speed: AI algorithms can process and analyze large volumes of user feedback data much faster than humans can.
- Accuracy: AI algorithms can identify patterns and insights in user feedback data that might be difficult for humans to detect.
- Consistency: AI algorithms can provide consistent results even with a large volume of data, avoiding human biases.
- Sentiment analysis: AI algorithms can classify user feedback data into positive, negative, or neutral categories, providing a more nuanced understanding of user feedback.

The use of AI in user feedback analysis has the potential to significantly improve the quality and efficiency of the process, helping organizations better understand their customers and make informed decisions.

### 4.2 Natural Language Processing (NLP) Contributions in User Feedback Analysis

Natural Language Processing (NLP) is a key technology that has made a significant contribution to user feedback analysis. NLP allows computers to understand and analyze human language, making it possible to extract insights from large volumes of user feedback data. Some of the key benefits of using NLP for user feedback analysis include:

- Text classification: NLP algorithms can classify user feedback data into different categories based on its content, making it easier to understand and analyze the data.
- Sentiment analysis: NLP algorithms can determine the sentiment behind the user feedback data, providing a more nuanced understanding of user opinions.
- Keyword extraction: NLP algorithms can extract important keywords and phrases from user feedback data, making it easier to identify common themes and trends.
- Topic modeling: NLP algorithms can identify the main topics being discussed in user feedback data, allowing organizations to understand what users are talking about and prioritize their responses.

The use of NLP in user feedback analysis can help organizations better understand their customers and make informed decision.

**5. Limitations**

This type of study presents several limitations that should be acknowledged. Firstly, the AI and NLP techniques applied are dependent on the quality and extent of the data available, which may affect the comprehensiveness of the user feedback analysis. Limitations in the algorithms' ability to understand context and sarcasm could also lead to potential inaccuracies in sentiment analysis. Additionally, the study focuses on a specific product type within a particular industry, which may limit the generalizability of the findings to other sectors or product categories. There may also be inherent biases in the user feedback collected, as it may not represent the views of the entire customer base. Future research could address these limitations by incorporating a broader data set, utilizing more advanced AI models capable of context recognition, and extending the analysis to multiple product types across various industries.

**6. Conclusion and Future Work**

In conclusion, user feedback analysis using AI and NLP techniques offers organizations a valuable means to gain direct insights into user perceptions, needs, and preferences. This knowledge can drive informed decision-making, guide product development, enhance customer satisfaction, and ultimately contribute to improved organizational behavior and performance. This paper gives an example on how to use Artificial Intelligence (AI) and Natural Language Processing (NLP) to do user feedback analysis to investigate unique business value of three heavy machine crane products by providing direct insights into how users perceive a product or service.

This type of analysis is based on the direct experiences and opinions of real users, which can provide valuable information for improving the product or addressing customer concerns. Additionally, in the future study, analysis methods of AI and NLP will help user feedback analysis to uncover trends and patterns in customer behavior, preferences, avoid more pain points that might not be apparent through other forms of data analysis.

While this paper has demonstrated the effectiveness of AI and NLP techniques in user feedback analysis for understanding user perceptions and improving organizational behavior, there are several avenues for future research in this domain. Here are some potential areas for future investigation:

**Advanced Sentiment Analysis**: Future studies could explore more sophisticated sentiment analysis techniques, such as aspect-based sentiment analysis, to gain a deeper understanding of specific aspects or features of a product or service that impact user satisfaction. This approach would provide more granular insights for targeted improvements.

**Deep Learning Approaches**: Investigating the application of deep learning algorithms, such as recurrent neural networks (RNNs) or transformers, in user feedback analysis could potentially improve the accuracy and efficiency of sentiment analysis and topic modeling. These approaches could handle complex language structures and capture nuanced user sentiments more effectively.

**Multimodal Analysis**: In addition to text data, incorporating other modalities such as images, videos, and audio into user feedback analysis could provide a more comprehensive understanding of user experiences. Analyzing visual and auditory cues alongside textual feedback could unveil additional insights and enhance the decision- making process.

**Longitudinal Studies**: Longitudinal studies tracking user feedback over an extended period can reveal evolving trends and patterns in user preferences and needs. By analyzing feedback data collected at different time points, organizations can identify changing customer expectations and make proactive adjustments to their products and strategies.

**Ethical Considerations**: As AI and NLP techniques continue to advance, it becomes increasingly important to address ethical considerations associated with user feedback analysis. Future studies could explore ethical frameworks and guidelines to ensure the responsible and fair use of user data while maintaining privacy and transparency.

By exploring these avenues for future research, organizations can further harness the power of AI and NLP in user feedback analysis to continuously enhance organizational behavior, drive innovation, and deliver exceptional customer experiences.